\documentclass{article}
\usepackage{arxiv}
\usepackage[utf8]{inputenc} % allow utf-8 input
\usepackage[T1]{fontenc}    % use 8-bit T1 fonts
\usepackage{hyperref}       % hyperlinks
\usepackage{url}            % simple URL typesetting
\usepackage{booktabs}       % professional-quality tables
\usepackage{amsfonts}       % blackboard math symbols
\usepackage{nicefrac}       % compact symbols for 1/2, etc.
\usepackage{microtype}      % microtypography
\usepackage{multicol}
\usepackage{graphicx}
\graphicspath{ {./images/} }

\title{Feasibility of Proof Of Authority as a Consensus Protocol Model}

\author{
 Shashank Joshi \\
 Department of Computer Science and Engineering\\
 SRM Institute Of Science And Technology\\
 Kattankulathur, Tamil Nadu – 603203, \\
 \texttt{sj8559@srmist.edu.in} \\

}

\begin{document}
\maketitle
\begin{abstract}
\textbf{Blockchain is a type of decentralized distributed network which acts as an immutable digital ledger. Despite the absence of any central governing authority to validate the blocks in the ledger, it is considered secure and immutable due to the consensus protocol among various nodes of the network.  A consensus algorithm is a mechanism that guarantees the reliability of the blockchain and helps all connected nodes or peers to reach common ground regarding the present state of the blockchain network thus an ideal consensus algorithm must be secure, reliable, and fast. There are several different algorithms to reach a consensus among the nodes thus this article seeks to test the practicality of Proof of Authority in the blockchain network as a consensus algorithm and its comparison with current mainstream consensus algorithms. }
\end{abstract}


\section{Introduction}

The Concept of blockchain and the use of consensus algorithm was initially proposed in 1982 \cite{1}, but a quantum leap was made with the introduction of Bitcoin, a peer-to-peer electronic cash system \cite{2} in 2009. The main idea behind blockchain was to create a decentralized and growing list of blocks or records which are integrated with the help of cryptography protocols. As blockchain functions as a decentralized immutable ledger which records the transaction in the real-time, it gives rise to a complexity to validate and verify the transaction and make the system fault-tolerant, this can be achieved with the help of a consensus algorithm which acts as an instrument using which nodes in the network can coordinate and come to a common consensus or single truth without a central authority, thus a consensus algorithm is an essential element of a blockchain network that maintains the integrity and security of these distributed computing systems. Many consensus algorithms developed with the time, for instance initially classical consensus protocols were the only known protocols that use all-to-all voting for obtaining single truth, then Satoshi proposed longest-chain consensus protocols after that, there were various variants of consensus protocols that came into existence like the snow family of consensus protocols \cite{3}. The primary goal of all these developments in consensus protocols is to make the blockchain network more secure and increase the transaction per second in the blockchain, as the properties of an ideal consensus algorithm are:

\begin{enumerate}
    \item Scalability: A consensus algorithm should maintain a smooth operation of the network by eliminating the possibility of slow processing time, system bloating, lags, etc.
    \item Integrity: The result generated by all the non faulty nodes must be indistinguishable.
    \item Termination: Every non faulty node should participate in deciding the output \cite{4}.
    \item Cooperative: A consensus protocol should aim to achieve the maximum agreement in the network \cite{4}.
    \item Egalitarian:A consensus protocol must give equal weightage to every vote in the network \cite{3}.
\end{enumerate}
As an immutable decentralized system blockchain’s nodes act as a cross-platform runtime that provides services both as host and server and they are required to share the data across all the nodes to reach a single truth to carry out transactions. No permissions are required to be a node in a public blockchain network thus there is no guarantee of integrity from the node thus a node can alter the transaction data and hence it creates a fork in the network. To avoid block tempering every node should agree on a single source of truth thus a consensus algorithm is required\cite{6}. Different sorts of blockchains have different application scenarios. Therefore, blockchain consensus algorithms should be appropriate and should suit the needs of specific applications \cite{5}. As the cryptocurrency space evolved many consensus algorithms like PoW(Proof of Work ), PoS (Proof of Stake), etc. are developed and become the mainstream consensus protocols, but there are some drawbacks of these consensus algorithms thus to reduce these flaws PoA is a feasible alternative consensus algorithm.

\section{Proof of Authority}
\label{sec:headings}
Proof of Authority is a regulated or permissioned consensus protocol, a family of BFT algorithms \cite{7} that aim to reach a consortium through authorized nodes or validators.
The validators are restricted to a fixed set of n authorized nodes called sealers \cite{9}. The working of this protocol can be summarized as: the validating machines generate the new blocks of transaction which are required to be added in the network. The mechanism of accepting and mining new blocks in the network depends upon the configuration of the respective blockchain network. In this protocol, the reputation of the sealer is at stake, thus making malicious behaviour very damaging to the reputation of the entity. Therefore, validating nodes of the blockchain network are responsible for the security of the network. The whole architecture of PoA depends upon:
\begin{enumerate}
    \item Liveliness of the Network: As the framework of PoA depends on the reputation of the sealer hence the liveliness of the network implies that the reputation system of the network does not provide a possibility to reach a state at which the bad behaviour of the sealer no longer damages their reputation \cite{10}.
    
    \item Tolerance: In a blockchain network with n sealers a maximum of p nodes where \begin{equation} p<\frac{n}2    \end{equation} can be dishonest or byzantine. Therefore to solve this byzantine consensus problem \cite{8} with n nodes where \begin{math}n-p\end{math} nodes are required to be honest and they all should come to a single source of truth i.e. a unique block in the network.
    
    \item Monotonicity: The reputation system associated with the PoA framework should not be monotonic in nature in any one direction as it will decrease the trust factor of the network \cite{10}.

    \item Visibility: The sealers in the network are required to prove and validate their actual identity and the reputation stats of all the sealers must be global and open to all the nodes in the blockchain network. 

    \item Difficulty in becoming a sealer: The process of becoming a sealer should be very selective and rigorous so that we can lower down any potential threat to the network \cite{3}.The process of selection should be transparent and equal for all.

\end{enumerate}
In a PoA based blockchain network, the validating node or sealers should be incentivised to retain their reputation or gain subsequently as this will refrain sealers from associating with any malicious activity on the network. The PoA consensus depends upon the mining rotation schema \cite{11} i.e. the responsibility of mining new blocks and keeping the network safe is fully dependent and distributed among the validatory nodes, hence the action of a sealer is strictly monitored by other sealers in the network.

\subsection{Comparison With Proof of Work}
Proof of Work (PoW) is another consensus model used in various blockchain networks and cryptocurrencies like bitcoin etc \cite{12}. According to this consensus model, an amount of work or computing power is required to mine new blocks in the blockchain network. This mechanism is designed to prevent any malicious activity in the network and make it less susceptible to a potential attack on the network. In PoW based protocols the mining nodes are required to compete and solve an unknown complex mathematical challenge and obtain a result less than the target value before adding a new block in the network. This computational intensive approach is the core safety mechanism in this protocol. As PoW is a computationally intensive approach it slows down the tps (transaction per second) of the network and the time lag in accepting the new block may lead another miner to find the optimum solution as a proof of work thus creating a fork in the network which increase the time taken to reach on a consensus \cite{5}. PoA on the other hand is less resource-consuming and has a far better throughput as compared to PoW due to the limited number of validatory nodes in the network thus improving the scalability of the system. PoA also provides an easy development and maintenance of decentralized applications and a centralized fashion decreases the number of forks, therefore, scaling down the number of possible attacks on the network.   

\subsection{Comparison With Proof of Stake}
Proof of Stake (PoS) is a type of consensus model which was initially developed as an energy-saving alternative to PoW. In PoS, the leaders are selected based on their stakes in the blockchain network, i.e., the number of digital tokens owned by the node in the network. In this model, the selection of the validatory node to add new blocks in the network is pseudorandom in nature. In this consensus model, instead of solving a complex mathematical problem, users who want to add new blocks in the network must lock a share of their stakes in the network. The probability \begin{math}p_i\end{math} of node i to be selected as a validatory node or leader among N participatory nodes is given by: \begin{equation} p_i=\frac{s_i}{\sum_{j=1}^{N}s_j}  \end{equation}
(Where \begin{math}s_i\end{math} is the stake of i th node) \cite{13}. PoA in contrast is a reputation-based consensus model that relies on a limited number of validating nodes for adding new blocks in the network. PoA gives more importance to the reputation of validating nodes rather than on their stakes in the network. Thus, improving the scalability and throughput of the network. PoA also presents a better solution for a permissioned blockchain network as compared to PoS. 

\section{Advantages}
\begin{itemize}
    \item PoA is less resource-consuming as compared to other consensus models as there is no direct competition among the nodes to add new blocks because of the limited number of validatory nodes in the network.
    \item This consensus model can achieve more efficiency and throughput relative to other consensus models due to its centralized reputation-based approach to reach a consensus.
    \item PoA based consensus model also solves the problem of market-based fluctuating processing fees by maintaining it constant throughout the network.
    \item Byzantine fault tolerance of PoA based consensus models is better than non-distributed protocols of a typical centralized system \cite{11}.

    \item Because of less number of forks generated in the network, the PoA based consensus protocols have better transaction rate as compared to PoS or PoW.
    \item The PoA based consensus model is well suited for the maintenance and development of DApps and is ascendable as compared to other consensus models\cite{14}.
\end{itemize}
\section{Limitations}
\begin{itemize}
    \item The PoA consensus model relinquishes some degree of decentralization because of its mechanism to reach a consensus using validatory nodes thus we can conclude PoA sacrifices decentralization in order to achieve high throughput and scalability \cite{7}.

    \item The identity of validating nodes is public across the network, thus knowing the validators’ identity can cause third-party manipulation and interference in the blockchain network.

    \item PoA consensus models’ centralized approach introduces censorship and blacklisting in the blockchain network. 

    \item Byzantine fault tolerance of PoA based consensus models is better than non-distributed protocols of a typical centralized system \cite{11}.

    \item PoA is a reputation-based consensus model, but if a validatory node in the network is intended to do malicious activities in the network, the threat of harming the reputation does not shield the network from malicious activities and third-party involvement.

    \item The PoA consensus model is prone to Sybil and cloning attacks \cite{15} in which one or two sealers can manipulate a large number of validatory nodes by forging multiple identities.

\end{itemize}
\section{Conclusion}
In this paper, there is a detailed analysis of the feasibility of proof of authority as a consensus algorithm in a blockchain network. The inherent features of PoA make it a potential alternative to overcome the shortcoming of PoW and PoS consensus protocols. PoA, as a consensus model, sacrifices decentralization to achieve high throughput and scalability, which makes it a suitable option for the DApps development and maintenance, private blockchains, and notary-based applications which require users to trust validators. On the other hand, it is still unproven for the public blockchain networks like cryptocurrencies, which give more importance to decentralization. Still, the cost-efficiency and performance of this construct make it a potential alternative for mainstream consensus protocols and an emerging blockchain solution for private or permissioned blockchain networks.
\bibliographystyle{unsrt}  
\bibliography{references}  

\end{document}